\documentclass{iau}
\usepackage{graphicx}

\title[GYRE: A New Oscillation Code] 
{GYRE: A New Open-Source Stellar Oscillation Code}

\author[Townsend, Teitler \& Paxton]   
{Rich Townsend$^1$, Seth Teitler$^1$ \and Bill Paxton$^2$}

\affiliation{%
$^1$Department of Astronomy, University of Wisconsin-Madison, Madison, WI 53706, USA\\
$^2$Kavli Institute for Theoretical Physics, University of California, Santa Barbara, CA 93106, USA}

\pubyear{2013}
\volume{301}  
\pagerange{1--2}
\jname{Precision Asteroseismology}
\editors{W. Chaplin, J. Guzik, G. Handler \& A. Pigulski}
\begin{document}

\maketitle

\begin{abstract}
We introduce GYRE, a new open-source stellar oscillation code which
solves the adiabatic/non-adiabatic pulsation equations using a novel
Magnus Multiple Shooting (MMS) numerical scheme. The code has a global
error scaling of up to 6th order in the grid spacing, and can
therefore achieve high accuracy with few grid points. It is moreover
robust and efficiently makes use of multiple processor cores and/or
nodes. We present an example calculation using GYRE, and discuss
recent work to integrate GYRE into the asteroseismic optimization
module of the MESA stellar evolution code.
\end{abstract}


Interpreting the wealth of new observations provided by \textit{MOST},
\textit{CoRoT} and \textit{Kepler} requires the theorist's analog to
the telescope: a stellar oscillation code which calculates the
eigenfrequency spectrum of an arbitrary input stellar model. Comparing
a calculated spectrum against a measured one provides a concrete
metric for evaluating a model, and therefore constitutes the bread and
butter of quantitative asteroseismology.

There's no shortage of oscillation codes available to the community;
the nine codes reviewed in \cite{Moy2008} are likely only a fraction
of those being used on a day-to-day basis. However, automated
asteroseismic optimization tools such as AMP (\cite[Metcalfe
  \etal\ 2009]{Met2009}) and MESA (\cite[Paxton \etal\ 2013]{Pax2013})
are placing ever-increasing demands on these codes. A code will
typically be executed hundreds or thousands of times during an
optimization run, and must therefore make efficient use of available
computational resources such as multi-processor hardware. The code
must be robust, running and producing sensible output without manual
intervention. The code must have an accuracy that matches or exceeds
the frequency precision now achievable by satellite missions. Finally,
it is preferable that the code address the various physical processes
that inevitably complicate calculations, such as non-adiabaticity,
rotation, and magnetic fields.
 
Currently, there are no publicly available oscillation codes which
address all of these requirements. This motivated us to develop
another code, `GYRE', which is built on a novel Magnus Multiple
Shooting (MMS) scheme for solving both adiabatic and non-adiabatic
pulsation problems. GYRE and the MMS scheme are described in detail in
a forthcoming paper (\cite[Townsend \& Teitler 2013]{TowTei2013}). The code
is written in standard-conforming Fortran 2008 with a modular
architecture that allows straightforward extension to handle
more-complicated problems. To leverage multiple processor cores and/or
cluster nodes it is parallelized using a combination of OpenMP and
MPI.

\begin{figure}[t]
\begin{center}
 \includegraphics[width=5.5in]{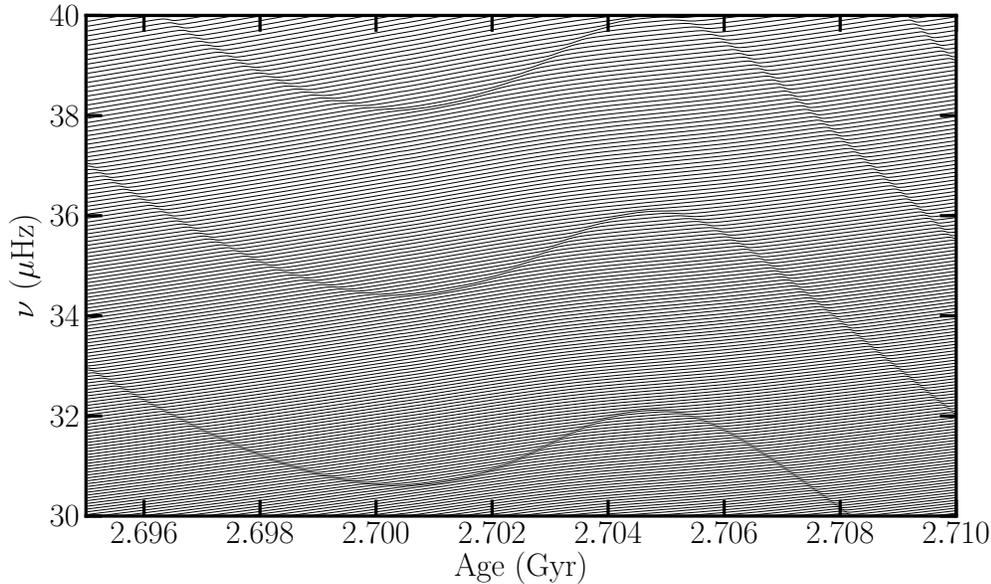}
\vspace*{-1.0 cm}
 \caption{Dipole-mode linear eigenfrequencies of a $1.5 \,M_{\odot}$
   MESA model, plotted as a function of stellar age during evolution
   through the RGB bump phase. Coupling between envelope p-modes and
   core g-modes is revealed in the avoided crossings; the reversal in
   the time evolution of the crossing frequencies, between 2.700 and
   2.705\,Gyr, arises from the temporary contraction of the stellar
   envelope during the bump phase.}
\label{fig:rgb_freq}
\end{center}
\end{figure}

As an illustration of GYRE in action, Fig.~\ref{fig:rgb_freq} presents
results from a simulation exploring how the dipole-mode oscillation
frequencies of a $1.5\,M_{\odot}$ MESA stellar model change as the
star evolves through the so-called red bump phase. (This phase occurs
when H-burning shell reaches the composition discontinuity left by the
convective envelope after first dredge-up, causing a temporary
reversal in the star's luminosity growth). The figure reveals that the
effects of bump passage can clearly be seen in the avoided crossings
which arise from coupling between core g-modes and envelope p-modes:
as the envelope contracts during the bump phase, the frequencies of
the avoided crossings correspondingly increase.

Since revision 5232 MESA includes GYRE as one of the oscillation codes
underpinning its asteroseismic optimization module (the other,
currently, is ADIPLS). Communication between MESA and GYRE is
accomplished through a simple application programming interface: MESA
passes a model to GYRE, which then returns a list of modes having
eigenfrequencies in a given range.

GYRE is open for use and distribution under the GNU General Public
License; our hope is that a community of practice will arise around
the code, bringing together users and developers to shape the
code's future evolution in ways that best serve the field
and its participants.  Source code, documentation and other materials
can be found at \texttt{http://www.astro.wisc.edu/\~{}townsend/gyre/}.

\acknowledgements
\noindent We acknowledge support from NSF awards AST-0908688 and AST-0904607 and
NASA award NNX12AC72G.

\begin{thebibliography}{}

\bibitem[Moya \etal\ (2008)]{Moy2008}
{Moya, A. \etal\ 2008, \textit{Ap\&SS}, 316, 231}

\bibitem[Metcalfe \etal (2009)]{Met2009}
{Metcalfe, T.~S., Creevey, O.~L., \& Christensen-Dalsgaard, J. 2009, \textit{ApJ}, 699, 373}

\bibitem[Paxton \etal (2013)]{Pax2013}
{Paxton, B. \etal\ 2013, \textit{ApJS}, 208, 4}

\bibitem[Townsend \& Teitler (2013)]{TowTei2013}
{Townsend, R.~H.~D., Teitler, S.~A. 2013, \textit{MNRAS}, in press (arXiv:1308.2965)}

\end{thebibliography}
\end{document}